\documentclass[prd,aps,
nofootinbib,%
twocolumn
]{revtex4-1}

\usepackage{graphicx}
\usepackage{amsmath}
\usepackage{amssymb}
\usepackage{color}
\usepackage{transparent}

\newcommand{\G}{\tilde{V}^{(\q)}}
\newcommand{\Bd}{{}^\star\!\tilde{B}}
\newcommand{\q}{\varepsilon}

\newcommand{\MKK}{M_{\rm KK}}

\newcommand{\rKK}{r_{\rm KK}}

\newcommand{\Nc}{N_{c}}

\newcommand{\ccsa}{b_1}
\newcommand{\ccsb}{b_2}

\newcommand{\cdbia}{\bar{b}_1}
\newcommand{\cdbib}{\bar{b}_2}
\newcommand{\cdbic}{\bar{b}_3}
\newcommand{\cdbid}{\bar{b}_4}
\newcommand{\cdbie}{\bar{b}_5}

\newcommand{\dbiC}{D}

\newcommand{\be}{\begin{equation}}
\newcommand{\ee}{\end{equation}}
\newcommand{\bea}{\begin{eqnarray}}
\newcommand{\eea}{\end{eqnarray}}
\newcommand{\nn}{\nonumber}
\newcommand{\6}{\partial }

\newcommand{\rkk}{r_\mathrm{KK}}

\begin{document}

\title{A broad pseudovector glueball from holographic QCD}

\author{Frederic Br\"unner, Josef Leutgeb, and Anton Rebhan}
\affiliation{Institute for Theoretical Physics, Vienna University of Technology (TU Wien)\\
        Wiedner Hauptstra\ss e 8-10, A-1040 Vienna, Austria}

\begin{abstract}
We study the decay of a  pseudovector glueball with quantum numbers ${J^{PC}}=1^{+-}$, the lightest glueball with spin different from zero or two, in the top-down holographic Witten-Sakai-Sugimoto model. 
This glueball is dual to the string-theoretic Kalb-Ramond tensor field, which completely fixes all its couplings by gauge invariance and $D$-brane anomaly inflow.
The dominant decay channels are determined by a Chern-Simons-like action for the flavor branes; couplings from the Dirac-Born-Infeld action turn out to be subdominant. 
While the resulting width is parametrically of order $N_c^{-2}\lambda^{-1}$ for $N_c$ colors and 't Hooft coupling $\lambda$,
when extrapolated to $N_c=3$ and with $\lambda$ matched to experimental data, the
prediction is that the pseudovector glueball is an extremely broad resonance.
\end{abstract}

\maketitle

\section{Introduction}

Glueballs are a cornerstone prediction of quantum chromodynamics (QCD)
\cite{Fritzsch:1972jv,*Fritzsch:1975tx,*Jaffe:1975fd}. Their spectrum
has been studied extensively in quenched lattice QCD 
up to and including spin $J=6$ \cite{Morningstar:1999rf,Chen:2005mg,Meyer:2004jc},
with results involving dynamical quarks becoming slowly available, too,
mostly showing only moderate effects from unquenching \cite{Gregory:2012hu}.
Experimentally, however, the situation remains unclear even for the lightest (scalar) glueball state,
where phenomenological models continue to be divided on which of the two isoscalar mesons $f_0(1500)$ and $f_0(1710)$ is
to be considered a glueball, even though the two mesons have distinctly different decay patterns, 
and to which extent there is mixing with $q\bar{q}$ states
\cite{Amsler:1995td,*Lee:1999kv,*Close:2001ga,*Amsler:2004ps,*Close:2005vf,*Giacosa:2005zt,*Albaladejo:2008qa,*Mathieu:2008me,Janowski:2014ppa,Cheng:2015iaa,Frere:2015xxa}.
The question of experimental candidates for the next lightest (tensor) glueball, for which lattice QCD
predicts a mass about roughly 2.4 GeV, is at least as unclear
\cite{Klempt:2007cp,Crede:2008vw}. For the pseudoscalar glueball no suitable candidate appears to be available
that would be in the mass range around 2.6 GeV predicted by (quenched) lattice QCD.
Of course the hope is that ongoing as well as future experiments, such as the PANDA experiment at FAIR \cite{Wiedner:2011mf},
will eventually provide fresh data and shed light on this long-standing
issue.

Unfortunately, information on experimental signatures such as width and decay patterns are not available 
from the mentioned lattice QCD studies, while phenomenological models typically come with a large number of
parameters and corresponding uncertainties.\footnote{Comparatively specific predictions for the decay patterns
of the pseudoscalar and vector glueball have been made, however, in Ref.~\cite{Eshraim:2012jv,*Eshraim:2016mds} and \cite{Giacosa:2016hrm}, respectively, albeit with undetermined total width.}

A new approach to the question of glueball decay properties has been pioneered in Ref.~\cite{Hashimoto:2007ze}
based on the top-down string-theoretic construction of a gauge/gravity dual to (large-$N_c$) QCD with
chiral quarks due to Witten \cite{Witten:1998zw} and Sakai and Sugimoto \cite{Sakai:2004cn},
which turns out to reproduce remarkably well, often quantitatively correct within 10--30\%, the known
features of low-energy QCD \cite{Sakai:2005yt}, while involving only one free dimensionless parameter -- the 't Hooft
coupling at the mass scale of the model.

{In particular, extrapolating this model to $N_c=3$ and setting mass scale and 't Hooft coupling 
such that low-energy QCD is matched \cite{Sakai:2004cn} yields a surprisingly good estimate of the hadronic decay rate of both $\rho$ and $\omega$ mesons, with
the decay of the latter being described by the Chern-Simons part of the flavor brane action \cite{Sakai:2005yt,Brunner:2015oqa}.}

{Motivated by this,}
in Ref.~\cite{Brunner:2015oqa} the calculation of glueball decay rates in the Witten-Sakai-Sugimoto (WSS) model has been revisited
for scalar and tensor glueballs. 
With finite quark masses included, the decay pattern of the predominantly dilatonic scalar glueball could be matched to
that of the experimental candidate $f_0(1710)$ \cite{1504.05815} if the glueball coupling to the mass term of
quark-antiquark mesons is such as to lead to small $\eta\eta'$ decay rates \cite{1510.07605}.
The lightest tensor glueball, on the other hand, is predicted to be a very broad state with $\Gamma/M \sim 0.5$
for $M\sim 2.4$ GeV \cite{Brunner:2015oqa,Rebhan:2016ecl} and may thereby have escaped experimental identification so far.

For the pseudoscalar glueball, however, a sharp state with very restricted decay patterns as well as production channels has been predicted in \cite{Brunner:2016ygk} within the WSS model, where this glueball is represented by a Ramond-Ramond one-form potential
with a central role in the realization of the U(1)$_\mathrm{A}$ anomaly.

In the present work, we consider the pseudovector ${J^{PC}}=1^{+-}$ glueball 
which is dual to the string-theoretic Kalb-Ramond two-form field, whose coupling to quark-antiquark
states in the WSS model is completely determined by gauge invariance and anomaly inflow arguments \cite{Green:1996dd}
at the leading order.

\section{The Kalb-Ramond field in the Witten-Sakai-Sugimoto model as pseudovector glueball}

An approximation to a holographic dual of low-energy large-$N_c$ Yang-Mills theory 
that is four-dimensional below the Kaluza-Klein
scale $M_{\mathrm{KK}}$ is given by Witten's model \cite{Witten:1998zw}, which is obtained as the near-horizon geometry of a stack of $N_c$ D4-branes compactified in its coordinate $x^4$ on a circle with radius $R_4=1/M_{\mathrm{KK}}$ with supersymmetry-breaking boundary conditions.
This can be obtained by dimensional reduction of an $\mathrm{AdS_7\times S^4}$ solution to eleven-dimensional supergravity with non-zero four-form flux $F_4$,
where $x^{11}$ is similarly compactified, with radius $R_{11}= g_s l_s$, where $g_s$ and $l_s=\sqrt{\alpha'}$ are string coupling and length, respectively.
The 11-dimensional line element is given by 
\bea
 ds^2&=&\frac{r^2}{L^2}\left[ f(r)dx_4^2+\eta_{\mu\nu}dx^\mu dx^\nu +dx_{11}^2\right]\nn\\
&&+\frac{L^2}{r^2}\frac{dr^2}{f(r)}+\frac{L^2}{4}d\Omega_4^2,
\eea
with $f(r)=1-\rkk^6/r^6$, where $L$ is the AdS radius, $r\in [\rkk,\infty)$ is the holographic coordinate bounded from below by $\rkk=M_{\mathrm{KK}}L^2/3$, and $d\Omega_4^2$ is the line element of a unit four-sphere. Greek indices indicate 4d Minkowski space with metric $\eta_{\mu\nu}$.

The spectrum of glueballs of the Witten model can be obtained by solving the equations of motion of the bosonic sector of eleven-dimensional supergravity linearized around the above background \cite{Brower:2000rp}. This yields towers of glueballs with quantum numbers $J^{PC}=0^{++},2^{++},0^{-+},1^{+-},1^{--}$
and masses proportional to $\MKK$. 
Pseudovector glueballs with $J^{PC}=1^{+-}$
correspond to fluctuations of the three-form field $A_3$ which after dimensional reduction yields the Kalb-Ramond two-form field $B_2$ of type-IIA string theory.
Inserting the ansatz
\bea
B_{\mu\nu}&=&A_{\mu\nu11}=r^3 N_4(r)\tilde B_{\mu\nu}(x^0,\mathbf{x}),\nn\\
&&A_{\rho 4 r}=6r^2N_4(r)\epsilon^{\alpha\beta\gamma\delta}\eta_{\rho\alpha} \frac{\6_\beta}{\Box}\tilde B_{\gamma\delta},
\eea
with $\eta^{\rho\mu}\6_\rho \tilde B_{\mu\nu}=0$
into the field equations of the 11-dimensional action
\begin{align}\label{L11}
2\kappa_{11}^{2}\mathcal{L}_{11}^{\left(b\right)} \supset
& 
-\frac{\sqrt{-g}}{2\cdot4!}F^{a_{1}...a_{4}}F_{a_{1}...a_{4}}\nonumber \\
& +\frac{\sqrt{g_{S^{4}}}}{2\cdot4!L}\epsilon^{a_{1}...a_{7}}A_{a_{1}...a_{3}}F_{a_{4}...a_{7}},
\end{align}
where 
$2\kappa_{11}^2=(2\pi)^8 l_s^{9} g_s^{3}$ and the indices $a_i$ refer to $\text{AdS}_7$ space, one obtains the mode equation \cite{Brower:2000rp}
\bea
&&\frac{d}{dr}r(r^6-\rkk^6)\frac{d}{dr}N_4(r)\nn\\
&&-(L^4 M^2 r^3-27r^5+\frac{9\rkk^6}{r})N_4(r)=0.
\eea
With the boundary condition $N'_4(\rkk)=0$ normalizable modes exist for a discrete mass spectrum with lowest value $M\approx 2.435 \MKK$.
Integrating (\ref{L11}) over the radial direction and the sphere $S^4$, the four-dimensional effective Lagrangian reads
\begin{align}\label{L4B2}
\mathcal{L}_{4}=-\frac{1}{4}\mathcal{C}_{B}\eta^{\rho\mu}\eta^{\sigma\nu}\tilde{B}_{\mu\nu}\left(M^{2}-\Box\right)\tilde{B}_{\rho\sigma}+\ldots
\end{align}
where 
\be
\mathcal{C}_{B}=R_{11}R_{4}L^{7}\frac{\pi^{4}}{3\kappa_{11}^{2}}\int\text{d}r\,r^{3}N_{4}(r)^{2}.
\ee
Setting $\mathcal{C}_{B}=1$, which yields the condition
\begin{equation}
N_{4}(\rkk)^{-1}  =0.00983838\,L^{3}\lambda^{\frac{1}{2}}N_{c}M_{\mathrm{KK}},
\end{equation}
corresponds to a canonical normalization of the Kalb-Ramond field, whose transverse polarizations
we  parametrize by
\be\label{Bpolarization}
\tilde{B}_{\mu\nu}=\frac{1}{\sqrt{\Box}}\eta^{\lambda\rho}\eta^{\kappa\sigma}\epsilon_{\mu\nu\lambda\kappa}\, \q_\rho\partial_\sigma \G(x),
\ee
where $\q_\rho$ is a unit spacelike vector. The hermitean field $\G(x)$ is then canonically normalized,
$\mathcal{L}_{4}=-\frac{1}{2}\G\left(M^{2}-\Box\right)\G+\ldots$.

Interactions with quark-antiquark states are introduced by using the construction of Sakai and Sugimoto, which places a stack of $N_f$ probe $\mathrm{D8\text{-}\overline{D8}}$-branes in the background described above, located at antipodal points on the $x^4$ circle \cite{Sakai:2004cn,Sakai:2005yt}. 
The nonabelian chiral symmetry $\mathrm{U}(N_f)_\mathrm{L}\times\mathrm{U}(N_f)_\mathrm{R}$ is broken spontaneously to $\mathrm{U}(N_f)_{\rm L+R}$ by the fact that the branes have to connect
in the bulk (at $r=\rKK$ for the antipodal embedding); the axial symmetry $\mathrm{U}(1)_\mathrm{A}$ is broken by the Witten-Veneziano mechanism \cite{Armoni:2004dc,Barbon:2004dq,Sakai:2004cn} .
The gauge fields $A_M$ living on the flavor branes are dual to mesonic states. Fluctuations of the background translate into fluctuations of the probe branes, and therefore effective interactions may be derived from the nine-dimensional worldvolume action of the latter. It consists of the Dirac-Born-Infeld (DBI) action%
\footnote{In the following we switch from the 11-dimensional metric to the 10-dimensional string frame metric.}
\begin{equation}\label{DBI}
 S_{\mathrm{DBI}}=-T_8\;\mathrm{STr}\int d^9 x\; e^{-\phi}\sqrt{-\mathrm{det}(g_{MN}+\mathcal{F}_{MN})},
\end{equation}
where $\mathrm{STr}$ denotes the symmetrized trace, and a Chern-Simons action involving the Ramond-Ramond fields $C_j$ 
of type-IIA string theory \cite{Green:1996dd}\footnote{Here we are dropping
the A-roof genus curvature contributions corresponding to mixed gauge-gravitational anomalies which
would give rise to a vertex between two tensor glueballs and a flavor singlet meson \cite{Anderson:2014jia}.}
\be
S_{\mathrm{CS}}=T_8\;\mathrm{Tr}\int e^{\mathcal{F}_2}\wedge\sum_j C_{2j+1},
\ee
with $T_8=(2\pi)^{-8} l_s^{-9}$, $\mathcal{F}_{MN}\equiv2\pi l_s^2F_{MN}+B_{MN}$, where $F_{MN}$ are the components of the $\mathrm{U}(N_f)$ field strength $F_2=dA_1-i A_1\wedge A_1$, and $B_{MN}$ the components of the Kalb-Ramond field $B_2$. Upon partial integration, 
the Chern-Simons action contains a term linear in $B_2$, 
which after integration over $S^4$ gives rise to the 5-dimensional integral
\be\label{CS}
S_{\mathrm{CS}}^{\left( B_2\right)} =\frac{1}{2}(2 \pi l_s^2)^2\;T_8\;L^3 \pi^2 g_s^{-1}\mathrm{Tr}\int A_1\wedge F_2\wedge B_2.
\ee
In order to derive an effective interaction Lagrangian we also need the explicit form of the fluctuations for the gauge field,
which for $N_f=3$ are given as usual \cite{Sakai:2004cn} by
\begin{equation}
 A_Z=\frac{\rkk^2}{2L} \phi_0(Z)\, \Pi^a T^a,\;\;\;\;\;\;A_\mu=\psi_1(Z)\,V_\mu^a T^a,
\end{equation}
with eight generators of $\mathrm{SU(3)}$ multiplying pseudoscalar and vector octets, $\Pi^a\equiv \Pi^a(x^\alpha)$ and $V_\mu^a\equiv V_\mu^a(x^\alpha)$,  satisfying $\mathrm{Tr}\,T^a T^b=\delta^{ab}$, $T^a=\frac{1}{\sqrt{2}}\lambda^a$ for $a\in\{1,\dots,8\}$; defining $T^0\equiv\frac{1}{\sqrt{3}}\mathbf{1}$ the singlet component of the pseudoscalar/vector meson nonet are included as $\Pi^0\equiv \eta_0$ and $V^0\equiv\omega_0$.
The coordinate $Z$ is defined by $Z^2=r^6/\rkk^6-1$ and parametrises the probe brane pairs along the holographic direction from $Z=-\infty$ to
$+\infty$. 
{(For the form and normalization of the mode functions $\phi_0$ and $\psi_1$ see e.g.\ Ref.~\cite{Brunner:2015oqa}.)}
In the chiral WSS model, the pseudoscalar octet states {(with mode function $\phi_0$)} are massless, 
whereas the vector mesons form a tower of massive states with {mode functions $\psi_n$} of alternating parity
beginning with $m_1=\sqrt{0.699}\MKK$ which sets the scale \cite{Sakai:2004cn} $\MKK=949$ MeV when $m_1$ is identified with the mass of $\rho$ and $\omega$ vector mesons.

Inserting these modes into Eq. (\ref{CS}) and integrating out all other dimensions, we obtain the four-dimensional interaction Lagrangian
\bea\label{int}
 \mathcal{L}_{4,\mathrm{int}}^{\mathrm{CS}} &\supset & {\ccsa}(\Pi^a\partial_\mu V_\nu^a+V_\mu^a\partial_\nu \Pi^a) \, \Bd^{\mu\nu} \nn\\
 &&-i{\ccsb}\mathrm{Tr}(T^a[T^b,T^c])\, \Pi^a V^b_\mu V^c_\nu \,\Bd^{\mu\nu},
\eea
where $\Bd^{\mu\nu}=\frac12 \epsilon^{\mu\nu\rho\sigma}\tilde{B}_{\rho\sigma}$,
\begin{equation}\label{coeff1}
 \ccsa=4C\int r^3\phi_0 \psi_1 N_4 dZ= 56.027 \, N_c^{-1} \lambda^{-\frac{1}{2}},
\end{equation}
and
\begin{equation}\label{coeff2}
 \ccsb=6C\int r^3\phi_0 \psi_1^2 N_4 dZ= 2571.72 \, N_c^{-\frac{3}{2}} \lambda^{-1},
\end{equation}
with $C=T_8 (2\pi l_s^2 \rKK L \pi)^2 /(16 g_s)$.

\begin{table}[t]
\begin{tabular}{cc}
\toprule
decay channel & $\Gamma/M$  \\
\colrule
$\pi\rho$ & 0.3624 \dots 0.4803 \\
$K K^*$ &  0.1945 \dots 0.2578 \\
$\eta \omega$ & 0.0530 \dots 0.0941 \\
$\eta \phi$ & 0.0086 \dots 0.0076 \\
$\eta' \omega$ & 0.0168 \dots 0.0203 \\
$\eta' \phi$ & 0.0020 \dots 0.0079 \\
\colrule
$\pi\rho\rho$ & 0.2595 \dots 0.4556\\
$\pi K^* K^* $ & 0.0213 \dots 0.0375 \\
$K K^*\rho$ & 0.0032 \dots 0.0056 \\
$K K^* \omega$ & 0.0011 \dots 0.0019 \\
\colrule
\emph{total} & \textbf{0.9225} \dots \textbf{1.3685}\\

\botrule
\end{tabular}
\caption{Results for the decay rates of a pseudoscalar glueball of mass $M=2311\;\mathrm{MeV}$ into pseudoscalar and vector mesons as determined by Eq. \eqref{int}.}
\label{results}
\end{table}

For an on-shell pseudovector glueball with polarization $\q$ one has $\Bd_{\mu\nu}=\frac1M(\q_\mu\6_\nu-\q_\nu\6_\mu)\G$.
The decay rates for $\G$ into various combinations of pseudoscalar and vector mesons may be evaluated by performing a phase space integral over the corresponding amplitudes and summing/averaging over polarisations of outgoing/ingoing particles. The results are summarised in Table \ref{results}, where we list dimensionless ratios of decay rates to the pseudovector glueball mass for all channels. The glueball mass is taken to be the WSS model result\footnote{%
If one were to extrapolate the $1^{+-}$ glueball mass to values up to $\sim 3$ GeV as indicated by lattice QCD studies \cite{Gregory:2012hu},
the decay width would increase significantly, by a factor of up to $\sim 2.5$, making it even much larger than the mass.}
$M\approx2311 \,\mathrm{MeV}$, $N_c=3$ and we interpolate  between two different fits for the 't Hooft coupling given by $\lambda=16.63\dots 12.55$. 
In the phase space integral we used experimental meson masses \cite{PDG18} and thus decays into some combinations of heavier vector mesons than the ones listed in the table are excluded kinematically. 

Additionally we have used the relations
\begin{align}\label{etamixing}
 \eta &= \eta_8 \cos \theta_P - \eta_0 \sin \theta_P\nonumber\\
 \eta' &= \eta_8 \sin \theta_P + \eta_0 \cos \theta_P, 
\end{align}
with the mixing angle resulting from combining bare quark masses (introduced e.g.\ through worldsheet instantons \cite{Aharony:2008an,*Hashimoto:2008sr,*McNees:2008km})
with the Witten-Veneziano mechanism \cite{Brunner:2016ygk}
\begin{equation}
 \theta_P=\frac{1}{2}\arctan\frac{2\sqrt{2}}{1-\frac{3}{2}m_0^2/(m_K^2-m_\pi^2)},
\end{equation}
where $m_0$ is the Witten-Veneziano mass of the WSS model that is given by \cite{Sakai:2004cn}
\begin{equation}
 m_0^2=\frac{N_f}{27\pi^2 N_c}\lambda^2 M_{\mathrm{KK}}^2.
\end{equation}
Moreover we have allowed for a mixing of $\omega$ and $\phi$, 
for simplicity with ideal mixing angle $\theta_V=
\arctan(1/\sqrt2)$ 
\cite{Amsler:2007cyq}.

In principle there are also numerous nonvanishing interaction vertices arising from the DBI action of Eq. \eqref{DBI}. However, as we show in Appendix \ref{appendix:DBI}, they are suppressed parametrically with order $N_c^{-\frac{3}{2}}\lambda^{-2}$ as well as by their numerical value as compared to the contributions from the Chern-Simons term. 
Besides vertices from terms linear in $B^{\mu\nu}$ (from higher-order terms in the DBI action), there is also a quadratic term $\propto B_{\mu\nu}B^{\mu\nu}$
from the lowest-order part of the DBI action which has the form of a correction to the mass term of the pseudovector glueball proportional to $\lambda^2 N_f/N_c$. 
As previously in Refs.~\cite{Hashimoto:2007ze,Brunner:2015oqa,Brunner:2016ygk,1504.05815,1510.07605}, such corrections to the masses of glueballs are neglected because the decay amplitude derived from the (probe) brane
action is itself a quantity of order $N_f/N_c$ so that the effects of these corrections on glueball decay rates are formally of higher order; the inclusion of such mass corrections would have to be considered together with backreactions
of the flavor branes on the supergravity background (e.g.\ along the lines of Ref.~\cite{Burrington:2007qd}), which is beyond the scope of the present paper.
(At any rate, as we discuss further in the Appendix, these terms quadratic in $B_{\mu\nu}$ turn out to be numerically small for our
choice of parameters and would only slightly increase our results for the decay rate.)

\section{Discussion}

In previous work \cite{1504.05815,1510.07605} we have found that the decay rate and the branching ratios of the predominantly dilatonic scalar glueball
calculated in the WSS model with finite quark masses added match remarkably well with existing data for the experimental scalar glueball candidates,
the isoscalar $f_0(1710)$, which some recent phenomenological models \cite{Janowski:2014ppa,Cheng:2015iaa} also favor as a meson with dominant glueball content.
The coupling of the tensor glueball to quark-antiquark states, which is parametrically of the same order, turns out, however, to lead to
a numerically large decay width with $\Gamma/M\sim 0.5$ for $M\sim 2.4$ GeV \cite{Brunner:2015oqa,Rebhan:2016ecl}, which is perhaps too large
to fit existing experimental candidates among $f_2$ mesons.
More recently, two of us have calculated the decay pattern of the next heavier glueball, a pseudoscalar, which is dual to the $C_1$ form field
that plays a central role for the $\mathrm{U}(1)_\mathrm{A}$ anomaly in the WSS model and the appearance of a Witten-Veneziano mass for $\eta$ and $\eta'$ mesons. 
Here the WSS model predicts a very narrow state \cite{Brunner:2016ygk}.

In the present work, we have considered the lightest glueball with spin different from zero or two,
a pseudovector which in the WSS model is dual to the Kalb-Ramond two-form field $B_{\mu\nu}$. This has direct couplings to quark-antiquark mesons
through the action of D8 branes, which are completely determined by gauge invariance and anomaly inflow arguments \cite{Green:1996dd}.
The latter prescribe the structure of the Chern-Simons part of the action that turns out to provide the leading contributions for the decay
amplitude of the pseudovector glueball. With the usual set of parameters of the WSS model, the decay width of this glueball turns out to
be extremely large {numerically}, $\Gamma/M \sim 1$ (or even larger when $M$ is extrapolated to the higher values indicated by lattice QCD),
{although parametrically the two-body and three-body decays are of order $N_c^{-2}\lambda^{-1}$ and $N_c^{-3}\lambda^{-2}$, respectively.}
{Like the tensor glueball, the pseudovector glueball is thus predicted to be a broad resonance, but even more so because of the 
large numerical values of the prefactors in the interaction vertices (\ref{coeff1}) and (\ref{coeff2}).}

It would also be interesting to consider the even heavier
vector glueball ${J^{PC}}=1^{--}$, which is carried by a different combination 
of the Kalb-Ramond field and the Ramond-Ramond 3-form field \cite{Brower:2000rp}, and to investigate whether it
is similarly unstable to leading order of the glueball decay calculations
in the WSS model.\footnote{Glueballs with spin above two are beyond the supergravity description and require
string-theoretic calculations within the WSS model \cite{Sonnenschein:2017ylo,*Peeters:2018xei}.}

{In the case of the hadronic decay width of $\rho$ and $\omega$ mesons, the prediction of the WSS model turns out to be surprisingly accurate
when extrapolated to $N_c=3$ \cite{Sakai:2005yt,Brunner:2015oqa}. Of course, this may cease to be the case for the more massive states
of tensor and pseudovector glueballs, in particular
when the parametrically small results for $\Gamma/M$ turn out to be numerically unsuppressed; 
so they should perhaps be just taken as a crude estimate. 
(After all, the WSS model appears to underestimate significantly the mass of the tensor and pseudovector glueball when compared to lattice results.)
Even so, the WSS result obtained here clearly indicates a broad pseudovector glueball and thus}
suggests that it would be very difficult to identify the pseudovector glueball in the hadron spectrum
(as {an isolated} non-$q\bar q$ isoscalar $h_1$ meson).

{While being an issue beyond the WSS model, where $h_1$ mesons from the $q\bar q$ spectrum
are not described by supergravity fields but only by string states,
mixing of the pseudovector glueball with other $h_1$ mesons may be non-negligible because of the large decay width of the former.
(A similar situation may be the case for tensor glueballs.)
The WSS results therefore also suggest that phenomenological models of $h_1$ (as well as $f_2$) mesons
should take into account possibly important glueball content of those states.
}

\begin{acknowledgments}
A.R. would like to thank Claude Amsler and Gunnar Bali for useful discussions.
This work was supported by the Austrian Science
Fund FWF, project no.\ P26366 and P30822, and the FWF doctoral program
Particles \& Interactions, project no.\ W1252.
\end{acknowledgments}

\appendix
\section{Pseudovector glueball couplings from the DBI action}
\label{appendix:DBI}
Here we calculate the coupling constants of the pseudo-vector glueball
to 3 mesons obtained from the DBI action
\begin{align}
S_{\mathrm{DBI}}= & -T_{8}\int d^{9}xe^{-\phi}\,\mathrm{STr}\,\sqrt{-\text{det}\left(g_{MN}+\mathcal{F}_{MN}\right)}\nn\\
= & -T_{8}\int d^{9}xe^{-\phi}\sqrt{-g}\,\mathrm{STr}\Bigl[1+\frac{1}{2}\text{tr}\Bigl(-\frac{1}{2}\left(g^{-1}\mathcal{F}\right)^{2}\nn\\
&+\frac{1}{3}\left(g^{-1}\mathcal{F}\right)^{3}-\frac{1}{4}\left(g^{-1}\mathcal{F}\right)^{4}\Bigr)\nn\\
 & +\frac{1}{32}\text{tr}\left(\left(g^{-1}\mathcal{F}\right)^{2}\right)^{2}\Bigr]+\mathcal{O}\left(\mathcal{F}^{5}\right),
\end{align}
where STr involves symmetrization of the U($N_f$) matrices and tr refers to spacetime indices.
Only the term of fourth order in $\mathcal{F}$ contains a vertex of a pseudovector glueball with quark-antiquark states.
In the absence of additional metric or dilaton fluctuations, the second-order (mixing) term vanishes because of transversality of $B_{\mu\nu}$.
The cubic term is also zero if at least one $\mathcal{F}$
is a singlet because $\text{Tr}\left(\text{tr}\left(\mathcal{F}\mathcal{F}\mathcal{F}^{0}\right)\right)=\text{tr}\left(T\mathcal{F}^{0}\right)$
with $T$ being the symmetric matrix $T_{MN}=\mathcal{F}_{MO}^{a}g^{OP}\mathcal{F}_{PN}^{a}$.

To estimate the importance of these vertices we set $N_{f}=1$ and
expand the relevant coupling terms
\bea\label{DBIint}
 \mathcal{L}_{4,\mathrm{int}}^{\mathrm{DBI}} &\supset & 
\cdbia{F}^{\nu\rho}{F}_{\rho\sigma}{F}^{\sigma\mu}\tilde{B}_{\mu\nu}+\cdbib{F}^{\rho\sigma}{F}_{\rho\sigma}{F}^{\mu\nu}\tilde{B}_{\mu\nu} \nn\\
&&\cdbic\partial^{\rho}\Pi\partial_{\rho}\Pi{F}^{\mu\nu}\tilde{B}_{\mu\nu}+\cdbid V^{\rho}\partial_{\rho}\Pi{F}^{\mu\nu}\tilde{B}_{\mu\nu}\nn\\
&&+\cdbie V^{\rho}V_{\rho}{F}^{\mu\nu}\tilde{B}_{\mu\nu},
\eea
where
$F^{\mu\nu}=\6^\mu V^\nu-\6^\nu V^\mu$
with coupling constants
\bea
\cdbia&= & 36 \dbiC \int dZ K^{-\frac{5}{6}} \psi_{1}^{3}N_{4} \nn\\
&=& 0.0000375351\,N_{c}^{-\frac{3}{2}}\lambda^{-2},\nn\\
\cdbib&= &  -{\cdbia}/{4},\nn\\
\cdbic&= &-\frac{\dbiC}{18} \MKK^6 L^6\int dZ K^{\frac{1}{2}} N_{4}\phi_{0}^2\psi_{1} \nn\\&=& -0.0000168453\,N_{c}^{-\frac{3}{2}}\lambda^{-2},\nn\\
\cdbid&= & 2\dbiC\MKK^4L^3\int dZ K^{\frac{1}{2}}N_{4}\phi_{0}\partial_{Z}\psi_{1} \nn\\&=& -0.00875721\,N_{c}^{-\frac{3}{2}}\lambda^{-2},\nn\\
\cdbie&= & -18\dbiC\MKK^2\int dZ  K^{\frac{1}{2}}N_{4} \psi_{1} \left(\partial_{Z}\psi_{1}\right)^2 \nn\\&=&
-1.78464\,N_{c}^{-\frac{3}{2}}\lambda^{-2},
\eea
where $\dbiC={L^3 \Nc}/({576 \MKK^2 \pi^2})$ and $K=1+Z^2  $.

These vertices, which couple a pseudovector glueball to three quark-antiquark mesons,
are suppressed by an extra factor $\lambda^{-1}$ compared to (\ref{coeff2}) as well as
by a much smaller numerical prefactor; they are therefore
negligibly small compared
to the vertices obtained from the CS action.

As mentioned in the text, besides linear terms in $B_{\mu\nu}$, the DBI action also contains
a term proportional to $B_{\mu\nu}B^{\mu\nu}$ which would appear as a correction arising from the flavor branes to the mass term
of the four-dimensional effective Lagrangian (\ref{L4B2}) resulting from the 10-dimensional supergravity action.
Explicitly, it reads
\bea\label{deltaM2}
\delta\mathcal{L}_4&=&-\frac{1}{4}\delta M^2\eta^{\rho\mu}\eta^{\sigma\nu}\tilde{B}_{\mu\nu}\tilde{B}_{\rho\sigma},\nn\\
&& \delta M^2 = 0.0035066 \frac{N_f}{N_c} \lambda^2 \MKK^2.
\eea
Taken at face value, this would mean an increase of the pseudovector glueball mass by 8\%, from 2311 to 2416\ldots2493 MeV,
for our choice of parameters. However, this contribution from the flavor branes is localized in the 10-dimensional bulk.
A complete calculation would require a corrected mode equation for the (bulk) Kalb-Ramond field on a 10-dimensional
background with order $N_f/N_c$ backreaction terms along the lines of Ref.~\cite{Burrington:2007qd}.
The numerical smallness of the partial result (\ref{deltaM2}) seems to indicate, however, that such an inclusion of backreaction
effects may be carried out in a perturbative manner even for parameters which correspond to extrapolations to QCD.

\bibliographystyle{JHEP}
\bibliography{glueballdecay}

\providecommand{\href}[2]{#2}\begingroup\raggedright\begin{thebibliography}{10}

\bibitem{Fritzsch:1972jv}
H.~Fritzsch and M.~Gell-Mann, {\it {Current algebra: Quarks and what else?}},
  {\em eConf} {\bf C720906V2} (1972) 135--165,
  [\href{http://arxiv.org/abs/hep-ph/0208010}{{\tt hep-ph/0208010}}].

\bibitem{Fritzsch:1975tx}
H.~Fritzsch and P.~Minkowski, {\it {$\Psi$ Resonances, Gluons and the Zweig
  Rule}},  {\em Nuovo Cim.} {\bf A30} (1975) 393.

\bibitem{Jaffe:1975fd}
R.~Jaffe and K.~Johnson, {\it {Unconventional States of Confined Quarks and
  Gluons}},  {\em Phys.Lett.} {\bf B60} (1976) 201.

\bibitem{Morningstar:1999rf}
C.~J. Morningstar and M.~J. Peardon, {\it {The Glueball spectrum from an
  anisotropic lattice study}},  {\em Phys.Rev.} {\bf D60} (1999) 034509,
  [\href{http://arxiv.org/abs/hep-lat/9901004}{{\tt hep-lat/9901004}}].

\bibitem{Chen:2005mg}
Y.~Chen, A.~Alexandru, S.~Dong, T.~Draper, I.~Horvath, et~al., {\it {Glueball
  spectrum and matrix elements on anisotropic lattices}},  {\em Phys.Rev.} {\bf
  D73} (2006) 014516, [\href{http://arxiv.org/abs/hep-lat/0510074}{{\tt
  hep-lat/0510074}}].

\bibitem{Meyer:2004jc}
H.~B. Meyer and M.~J. Teper, {\it {Glueball Regge trajectories and the pomeron:
  A Lattice study}},  {\em Phys. Lett.} {\bf B605} (2005) 344--354,
  [\href{http://arxiv.org/abs/hep-ph/0409183}{{\tt hep-ph/0409183}}].

\bibitem{Gregory:2012hu}
E.~Gregory, A.~Irving, B.~Lucini, C.~McNeile, A.~Rago, et~al., {\it {Towards
  the glueball spectrum from unquenched lattice QCD}},  {\em JHEP} {\bf 1210}
  (2012) 170, [\href{http://arxiv.org/abs/1208.1858}{{\tt arXiv:1208.1858}}].

\bibitem{Amsler:1995td}
C.~Amsler and F.~E. Close, {\it {Is $f_0(1500)$ a scalar glueball?}},  {\em
  Phys.Rev.} {\bf D53} (1996) 295--311,
  [\href{http://arxiv.org/abs/hep-ph/9507326}{{\tt hep-ph/9507326}}].

\bibitem{Lee:1999kv}
W.-J. Lee and D.~Weingarten, {\it {Scalar quarkonium masses and mixing with the
  lightest scalar glueball}},  {\em Phys.Rev.} {\bf D61} (1999) 014015,
  [\href{http://arxiv.org/abs/hep-lat/9910008}{{\tt hep-lat/9910008}}].

\bibitem{Close:2001ga}
F.~E. Close and A.~Kirk, {\it {Scalar glueball $q \bar q$ mixing above 1 GeV
  and implications for lattice QCD}},  {\em Eur.Phys.J.} {\bf C21} (2001)
  531--543, [\href{http://arxiv.org/abs/hep-ph/0103173}{{\tt hep-ph/0103173}}].

\bibitem{Amsler:2004ps}
C.~Amsler and N.~T{\"o}rnqvist, {\it {Mesons beyond the naive quark model}},
  {\em Phys.Rept.} {\bf 389} (2004) 61--117.

\bibitem{Close:2005vf}
F.~E. Close and Q.~Zhao, {\it {Production of $f_0(1710)$, $f_0(1500)$, and
  $f_0(1370)$ in $J/\psi$ hadronic decays}},  {\em Phys.Rev.} {\bf D71} (2005)
  094022, [\href{http://arxiv.org/abs/hep-ph/0504043}{{\tt hep-ph/0504043}}].

\bibitem{Giacosa:2005zt}
F.~Giacosa, T.~Gutsche, V.~Lyubovitskij, and A.~Faessler, {\it {Scalar nonet
  quarkonia and the scalar glueball: Mixing and decays in an effective chiral
  approach}},  {\em Phys.Rev.} {\bf D72} (2005) 094006,
  [\href{http://arxiv.org/abs/hep-ph/0509247}{{\tt hep-ph/0509247}}].

\bibitem{Albaladejo:2008qa}
M.~Albaladejo and J.~Oller, {\it {Identification of a Scalar Glueball}},  {\em
  Phys.Rev.Lett.} {\bf 101} (2008) 252002,
  [\href{http://arxiv.org/abs/0801.4929}{{\tt arXiv:0801.4929}}].

\bibitem{Mathieu:2008me}
V.~Mathieu, N.~Kochelev, and V.~Vento, {\it {The Physics of Glueballs}},  {\em
  Int.J.Mod.Phys.} {\bf E18} (2009) 1--49,
  [\href{http://arxiv.org/abs/0810.4453}{{\tt arXiv:0810.4453}}].

\bibitem{Janowski:2014ppa}
S.~Janowski, F.~Giacosa, and D.~H. Rischke, {\it {Is $f_0(1710)$ a glueball?}},
   {\em Phys.Rev.} {\bf D90} (2014) 114005,
  [\href{http://arxiv.org/abs/1408.4921}{{\tt arXiv:1408.4921}}].

\bibitem{Cheng:2015iaa}
H.-Y. Cheng, C.-K. Chua, and K.-F. Liu, {\it {Revisiting Scalar Glueballs}},
  {\em Phys. Rev.} {\bf D92} (2015) 094006,
  [\href{http://arxiv.org/abs/1503.06827}{{\tt arXiv:1503.06827}}].

\bibitem{Frere:2015xxa}
J.-M. Fr{\`e}re and J.~Heeck, {\it {Scalar glueballs: Constraints from the
  decays into $\eta$ or $\eta'$}},  {\em Phys. Rev.} {\bf D92} (2015) 114035,
  [\href{http://arxiv.org/abs/1506.04766}{{\tt arXiv:1506.04766}}].

\bibitem{Klempt:2007cp}
E.~Klempt and A.~Zaitsev, {\it {Glueballs, Hybrids, Multiquarks. Experimental
  facts versus QCD inspired concepts}},  {\em Phys.Rept.} {\bf 454} (2007)
  1--202, [\href{http://arxiv.org/abs/0708.4016}{{\tt arXiv:0708.4016}}].

\bibitem{Crede:2008vw}
V.~Crede and C.~Meyer, {\it {The Experimental Status of Glueballs}},  {\em
  Prog.Part.Nucl.Phys.} {\bf 63} (2009) 74--116,
  [\href{http://arxiv.org/abs/0812.0600}{{\tt arXiv:0812.0600}}].

\bibitem{Wiedner:2011mf}
U.~Wiedner, {\it {Future Prospects for Hadron Physics at PANDA}},  {\em
  Prog.Part.Nucl.Phys.} {\bf 66} (2011) 477--518,
  [\href{http://arxiv.org/abs/1104.3961}{{\tt arXiv:1104.3961}}].

\bibitem{Eshraim:2012jv}
W.~I. Eshraim, S.~Janowski, F.~Giacosa, and D.~H. Rischke, {\it {Decay of the
  pseudoscalar glueball into scalar and pseudoscalar mesons}},  {\em Phys.
  Rev.} {\bf D87} (2013) 054036, [\href{http://arxiv.org/abs/1208.6474}{{\tt
  arXiv:1208.6474}}].

\bibitem{Eshraim:2016mds}
W.~I. Eshraim and S.~Schramm, {\it {Decay modes of the excited pseudoscalar
  glueball}},  {\em Phys. Rev.} {\bf D95} (2017) 014028,
  [\href{http://arxiv.org/abs/1606.02207}{{\tt arXiv:1606.02207}}].

\bibitem{Giacosa:2016hrm}
F.~Giacosa, J.~Sammet, and S.~Janowski, {\it {Decays of the vector glueball}},
  {\em Phys. Rev.} {\bf D95} (2017) 114004,
  [\href{http://arxiv.org/abs/1607.03640}{{\tt arXiv:1607.03640}}].

\bibitem{Hashimoto:2007ze}
K.~Hashimoto, C.-I. Tan, and S.~Terashima, {\it {Glueball decay in holographic
  QCD}},  {\em Phys.Rev.} {\bf D77} (2008) 086001,
  [\href{http://arxiv.org/abs/0709.2208}{{\tt arXiv:0709.2208}}].

\bibitem{Witten:1998zw}
E.~Witten, {\it {Anti-de Sitter space, thermal phase transition, and
  confinement in gauge theories}},  {\em Adv.Theor.Math.Phys.} {\bf 2} (1998)
  505--532, [\href{http://arxiv.org/abs/hep-th/9803131}{{\tt hep-th/9803131}}].

\bibitem{Sakai:2004cn}
T.~Sakai and S.~Sugimoto, {\it {Low energy hadron physics in holographic QCD}},
   {\em Prog.Theor.Phys.} {\bf 113} (2005) 843--882,
  [\href{http://arxiv.org/abs/hep-th/0412141}{{\tt hep-th/0412141}}].

\bibitem{Sakai:2005yt}
T.~Sakai and S.~Sugimoto, {\it {More on a holographic dual of QCD}},  {\em
  Prog.Theor.Phys.} {\bf 114} (2005) 1083--1118,
  [\href{http://arxiv.org/abs/hep-th/0507073}{{\tt hep-th/0507073}}].

\bibitem{Brunner:2015oqa}
F.~Br{\"u}nner, D.~Parganlija, and A.~Rebhan, {\it {Glueball Decay Rates in the
  Witten-Sakai-Sugimoto Model}},  {\em Phys. Rev.} {\bf D91} (2015) 106002,
  [\href{http://arxiv.org/abs/1501.07906}{{\tt arXiv:1501.07906}}]. [Erratum:
  Phys. Rev.D93, 109903 (2016)].

\bibitem{1504.05815}
F.~Br{\"u}nner and A.~Rebhan, {\it {Nonchiral enhancement of scalar glueball
  decay in the Witten-Sakai-Sugimoto model}},  {\em Phys. Rev. Lett.} {\bf 115}
  (2015) 131601, [\href{http://arxiv.org/abs/1504.05815}{{\tt
  arXiv:1504.05815}}].

\bibitem{1510.07605}
F.~Br{\"u}nner and A.~Rebhan, {\it {Constraints on the $\eta\eta'$ decay rate
  of a scalar glueball from gauge/gravity duality}},  {\em Phys. Rev.} {\bf
  D92} (2015) 121902, [\href{http://arxiv.org/abs/1510.07605}{{\tt
  arXiv:1510.07605}}].

\bibitem{Rebhan:2016ecl}
A.~Rebhan, {\it {Scalar and tensor glueball decay rates from the
  Witten-Sakai-Sugimoto model}},  in {\em {Proceedings, 51st Rencontres de
  Moriond on QCD and High Energy Interactions: La Thuile, Italy, March 19-26,
  2016}}, pp.~81--84, 2016.
\newblock \href{http://arxiv.org/abs/1605.09028}{{\tt arXiv:1605.09028}}.

\bibitem{Brunner:2016ygk}
F.~Br{\"u}nner and A.~Rebhan, {\it {Holographic QCD predictions for production
  and decay of pseudoscalar glueballs}},  {\em Phys. Lett.} {\bf B770} (2017)
  124--130, [\href{http://arxiv.org/abs/1610.10034}{{\tt arXiv:1610.10034}}].

\bibitem{Green:1996dd}
M.~B. Green, J.~A. Harvey, and G.~W. Moore, {\it {I-brane inflow and anomalous
  couplings on D-branes}},  {\em Class. Quant. Grav.} {\bf 14} (1997) 47--52,
  [\href{http://arxiv.org/abs/hep-th/9605033}{{\tt hep-th/9605033}}].

\bibitem{Brower:2000rp}
R.~C. Brower, S.~D. Mathur, and C.-I. Tan, {\it {Glueball spectrum for QCD from
  AdS supergravity duality}},  {\em Nucl.Phys.} {\bf B587} (2000) 249--276,
  [\href{http://arxiv.org/abs/hep-th/0003115}{{\tt hep-th/0003115}}].

\bibitem{Armoni:2004dc}
A.~Armoni, {\it {Witten-Veneziano from Green-Schwarz}},  {\em JHEP} {\bf 0406}
  (2004) 019, [\href{http://arxiv.org/abs/hep-th/0404248}{{\tt
  hep-th/0404248}}].

\bibitem{Barbon:2004dq}
J.~L. Barbon, C.~Hoyos-Badajoz, D.~Mateos, and R.~C. Myers, {\it {The
  Holographic life of the $\eta'$}},  {\em JHEP} {\bf 0410} (2004) 029,
  [\href{http://arxiv.org/abs/hep-th/0404260}{{\tt hep-th/0404260}}].

\bibitem{Anderson:2014jia}
N.~Anderson, S.~K. Domokos, J.~A. Harvey, and N.~Mann, {\it {Central production
  of $\eta$ and $\eta'$ via double Pomeron exchange in the Sakai-Sugimoto
  model}},  {\em Phys. Rev.} {\bf D90} (2014) 086010,
  [\href{http://arxiv.org/abs/1406.7010}{{\tt arXiv:1406.7010}}].

\bibitem{PDG18}
M.~Tanabashi et~al., {\it {Review of Particle Physics}},  {\em Phys. Rev. D}
  {\bf 98} (2018) 030001.

\bibitem{Aharony:2008an}
O.~Aharony and D.~Kutasov, {\it {Holographic Duals of Long Open Strings}},
  {\em Phys.Rev.} {\bf D78} (2008) 026005,
  [\href{http://arxiv.org/abs/0803.3547}{{\tt arXiv:0803.3547}}].

\bibitem{Hashimoto:2008sr}
K.~Hashimoto, T.~Hirayama, F.-L. Lin, and H.-U. Yee, {\it {Quark Mass
  Deformation of Holographic Massless QCD}},  {\em JHEP} {\bf 0807} (2008) 089,
  [\href{http://arxiv.org/abs/0803.4192}{{\tt arXiv:0803.4192}}].

\bibitem{McNees:2008km}
R.~McNees, R.~C. Myers, and A.~Sinha, {\it {On quark masses in holographic
  QCD}},  {\em JHEP} {\bf 0811} (2008) 056,
  [\href{http://arxiv.org/abs/0807.5127}{{\tt arXiv:0807.5127}}].

\bibitem{Amsler:2007cyq}
C.~Amsler, {\em {Nuclear and particle physics}}.
\newblock IOPP, 2015.
\newblock DOI:10.1088/978-0-7503-1140-3.

\bibitem{Burrington:2007qd}
B.~A. Burrington, V.~S. Kaplunovsky, and J.~Sonnenschein, {\it {Localized
  Backreacted Flavor Branes in Holographic QCD}},  {\em JHEP} {\bf 0802} (2008)
  001, [\href{http://arxiv.org/abs/0708.1234}{{\tt arXiv:0708.1234}}].

\bibitem{Sonnenschein:2017ylo}
J.~Sonnenschein and D.~Weissman, {\it {The decay width of stringy hadrons}},
  {\em Nucl. Phys.} {\bf B927} (2018) 368--454,
  [\href{http://arxiv.org/abs/1705.10329}{{\tt arXiv:1705.10329}}].

\bibitem{Peeters:2018xei}
K.~Peeters, M.~Matuszewski, and M.~Zamaklar, {\it {Holographic meson decays via
  worldsheet instantons}},  {\em JHEP} {\bf 06} (2018) 083,
  [\href{http://arxiv.org/abs/1803.06318}{{\tt arXiv:1803.06318}}].

\end{thebibliography}\endgroup

\end{document}